\documentclass[a4paper,fleqn,usenatbib]{mnras}

\usepackage[T1]{fontenc}
\usepackage{ae,aecompl}

\usepackage{graphicx}	
\usepackage{amsmath}	
\usepackage{amssymb}	

\usepackage{newtxtext,newtxmath}

\usepackage{cuted}

\usepackage{comment}

\title[Polarisation spectra in molecular clouds]{
On the origin of V-shaped polarisation spectra in molecular clouds
}
  \author[D. Seifried et al. ]
  {D.~Seifried,$^1$\thanks{seifried@ph1.uni-koeln.de} S.~Walch,$^1$ T. Balduin$^{2}$
 \\
  $^1$Universit\"at zu K\"oln, I. Physikalisches Institut, Z\"ulpicher Str. 77, 50937 K\"oln, Germany\\
  $^2$University of Copenhagen, Niels Bohr Institute, Blegdamsvej 17-21, 2100 Copenhagen, Denmark\\
  }

\date{Released 2025}

\pubyear{2025}

\begin{document}
\label{firstpage}
\pagerange{\pageref{firstpage}--\pageref{lastpage}}
\maketitle

\begin{abstract}
We extend previous theoretical works to gain a better understanding of the origin of observed polarisation degree spectra of molecular clouds, which show a so-called V-shape, i.e. a pronounced minimum around 350~$\mu$m. For this purpose, we present results of two-phase dust models investigated with POLARIS. We also provide a guideline to calculate individual dust temperatures for different grain types in POLARIS. We show that V-shaped polarisation spectra can only be obtained if two dust phases, one dense and cold as well as one warm and dilute phase, are present along the line of sight. We find that the V-shape is the stronger pronounced the larger the density and temperature contrast between both phases is. In contrast to previous results, no correlation between the alignment efficiency of silicate grains and the dust temperature is required; carbonaceous grains are in general assumed to be not aligned with the magnetic field. By matching our model results with actual observations of V-shaped polarisation spectra, we show that in UV-illuminated regions (here the warm and dilute phase) carbon grain destruction might take place. This leads to a more pronounced V-shape with a minimum around 300~$\mu$m. In addition, we show that the dust spectral index and temperature of silicate grains affects the steepness of the polarisation spectrum at long wavelengths. Finally, we present a first polarisation spectrum obtained from a 3D, magneto-hydrodynamical molecular cloud simulation. It shows a flattening or even weakly pronounced minimum around 350~$\mu$m demonstrating the potential of such complex 3D simulations to study polarisation spectra.
\end{abstract}




\section{Introduction}
\label{sec:intro}

The polarised radiation emitted by dust grains in the interstellar medium (ISM) has attracted attention since its discovery in 1949 \citep{Hall49,Hiltner49}. In general, this polarisation is caused by dust grains aligning with their short axis parallel to the local magnetic field. Hence, studying the (degree of) polarisation allows to infer properties about the emitting dust grains and the interstellar magnetic field. A prominent example is the finding that the silicate features at 9.8 and 18.5~$\mu$m are strongly polarised \citep[e.g.][]{Aitken88,Smith00,Wright02,Whittet11}, whereas the corresponding carbon feature at 3.4~$\mu$m (C-H stretching mode) shows no or only a weak polarisation \citep{Adamson99,Chiar06}. This lead to the conclusion that carbonaceous grains are most likely not aligned with the interstellar magnetic field whereas silicate grains are.

The currently accepted paradigm is that the grain alignment is caused by radiative torques \citep[][but see also the review by \citealt{Ngoc22}]{Dolginov76,Draine96,Lazarian07,Hoang09}: the generally anisotropic interstellar radiation field spins up the dust grains. In consequence, the resulting magnetic dipole, which tends to be aligned with the short axis of the grain, causes the grain to precess with its shortest axis (associated with the largest moment of inertia) around the magnetic field. This in turn causes a preferred emission of polarised radiation perpendicular to the field lines.

For dense and cold regions in the ISM like molecular clouds, dust (polarisation) observations are usually taken in the far-infrared and sub-mm/mm range at wavelengths longer than several 10 $\mu$m up to a few mm. The analysis of the polarisation pattern, e.g. the relative orientation of field lines and density structures \citep[e.g.][see also \citealt{Pattle23} for a review]{PlanckXXXII,Soler17b,Jow18} can tell us about the relative importance of magnetic fields with respect to gravity \citep{Soler17,Seifried20a}. More information about the physics and spatial structure of the gas and the magnetic field might be encoded in the change of the polarisation degree (at a fixed wavelength) with e.g. the column density or the intensity \citep[e.g.][]{Fissel16,Pattle19,Ngoc21}. More recently, \citet{Hoang19a} and \citet{Hoang19b} suggested the radiative torque induced disruption (RAT-D) of grains. Its effect and thus information about the spinning up of dust grains can be inferred from the change of the polarisation degree with the line-of-sight averaged dust temperature  \citep[e.g.][but see also the review by \citealt{Tram22}]{Ngoc21,Tram21,Hoang22} 

Information about the dust properties in molecular clouds might also be inferred from the change of the polarisation degree with wavelength, the so-called polarisation spectrum. Over the last years a number of polarisation observations at different wavelengths towards various molecular clouds have been performed. The obtained polarisation spectra typically show large variations in their shapes, ranging from flat spectra \citep{PlanckXXII,Gandilo16,Ashton18,Shariff19,Cox25} over spectra either increasing or decreasing with wavelength \citep{Hildebrand99,Zeng13,Santos19,Cox25}.  Even within an individual source, SOFIA observations show spatial variations in the spectrum at wavelengths $\leq 214 \mu$m \citep{Santos19,Cox25}.

A particularly peculiar spectral type is the so-called \textit{V-shaped spectrum}, i.e. a spectrum with a pronounced minimum around 350~$\mu$m \citep[][and reference therein]{Hildebrand00,Vaillancourt07SPIE,Vaillancourt08,Vaillancourt12}. This V-shaped polarisation spectrum has just recently also been observed for OMC~1 with high-resolution observations \citep{Tram24}. In addition, a V-shaped spectrum has also been observed towards entire starburst galaxies \citep{Lopez22}. However, it is fair to note that some concerns related to systemic uncertainties in the observations have been raised concerning this increase of the polarisation degree at long wavelengths \citep{Cox25}.

Various effects, which have been suggested to influence the shape of the polarisation spectrum, are discussed in these publications. These include varying dust densities and temperatures along the line of sight as well as variations in the dust properties. Some of the seminal theoretical considerations date back to \citet{Hildebrand99,Hildebrand00}. The authors suggest that, in order to obtain a V-shaped polarisation spectrum, various dust components with different temperatures and different alignment efficiencies are required to exist along the line of sight. Specifically, an anticorrelation between the dust temperature and the alignment efficiency is suggested by the authors. Furthermore, some first results combining 3D, magneto-hydrodynamical (MHD) simulations with a dust polarisation radiative transfer were presented by \citet{Bethell07}, finding flat polarisation spectra longwards of 300~$\mu$m, with a steep decline at shorter wavelengths. 

Overall, a general consensus about how these polarisation spectra are shaped and what we can learn from them about specific dust properties and environmental conditions is still missing. Given the recent re-confirmation of particularly the V-shaped polarisation spectrum by \citet{Tram24}, this emphasises the need for a thorough investigation of the origin of such spectral shapes. For this reason, in this work we extend the previous theoretical works in a more quantitative manner. For this purpose, in Section~\ref{sec:numerics} we first present the basics of a simply toy model and a 3D simulation used in this work. In Section~\ref{sec:onephase} we present the results of the most simple toy model and discuss the impact of the dust spectral index. Next, in Section~\ref{sec:twophase} we extend the model to match recent observations by incorporating regions with different dust temperatures. We also discuss the effect of carbonaceous grain destruction on the polarisation spectrum. In Section~\ref{sec:discussion} we compare our results with recent observations and other theoretical works. We also give an outlook of how to extend this work to more sophisticated 3D, MHD simulations of molecular clouds. We conclude our work in Section~\ref{sec:conclusion}.

\section{Numerical models}
\label{sec:numerics}

In the following we discuss results obtained with the POLARIS code \citep{Reissl16} using a toy model to investigate the shape of the polarisation spectrum and the effect of various parameters on it. This toy model was introduced first by  \citet{Hildebrand99} in a semi-analytic fashion and also used in a subsequent work of \citet{Ashton18}. In order to provide a coherent overview, we first explain the underlying concepts of the toy model before we provide the details of its implementation in POLARIS. Towards the end of the section, we also discuss an extension to 3D, MHD simulations.

\subsection{The toy model: basic concepts}
\label{sec:theory}

The toy model used in this work aims to mimic the multiphase nature of the interstellar medium, specifically towards molecular clouds, in a simplified manner. For this purpose it contains different, (spatially) distinct phases. A single phase consists of a mix of silicate and carbonaceous grains with a fixed abundance ratio. Furthermore, within one phase the temperature distribution of both types of grains does not change, the two distributions can, however, differ from each other. 

\subsubsection{The one-phase case}
\label{sec:1phasetheory}

For the carbon- and silicate-based dust used in each phase we use the common dust size distribution given by \citet{Mathis77} of $n \propto a^{-3.5}$, where $n$ is the dust grain density of grains with radius $a$, with typical values for the lower and upper cut-off radius of \mbox{$a_\textrm{min}$ = 5~nm} and \mbox{$a_\textrm{max}$ = 500~nm}, respectively. The mass fraction of both grain types $i$ is denoted by $f_i$. We assume that $f_\textrm{sil}$ = 62.5\% of the dust mass is contained in silicate grains, and $f_\textrm{carb}$ = 37.5\% in carbonaceous grains. We note that from hereon, we use the subscripts ``sil'' and ``carb'' when referring to silicate and carbonaceous grains, respectively. As mentioned in Section~\ref{sec:intro}, we assume that carbonaceous grains are not aligned with the magnetic field and thus do not contribute to the overall polarised emission. Both grain types can have individual temperatures, $T_\textrm{sil}(a)$ and $T_\textrm{carb}(a)$, which in addition can be size-dependent. 

The normalised absorption cross section of each grain is size- and wavelength-dependent and is in our case given by the database provided by POLARIS. A common parametrisation is given by \citet{Draine11} as
\begin{equation}
Q_{\textrm{abs,} i}(a,\lambda) = Q_{0, i} \times \left( \frac{a}{\textrm{1 \, $\mu$ m}} \right)  \times \left( \frac{\lambda}{\textrm{100 \, $\mu$ m}}\right)^{-\beta_i(\lambda)} \, ,
\label{eq:Qabs}
\end{equation}
where $i$ indicates the dust species. Here, $\lambda$ is the considered wavelength, $\beta(\lambda)$ the wavelength-dependent spectral index, which typically has values between 1.6 and 2, and $Q_0$ the normalisation factor.

With this, the resulting total intensity is
\begin{equation}
I_\textrm{1-phase} = \sum_{i} f_i  \times \int_{a_\textrm{min}}^{a_\textrm{max}} B_{\nu}(T_i(a)) \times \pi a^2 \times Q_{\textrm{abs,} i}(a,\lambda) \times a^{-3.5} \textrm{d}a     \, ,
\label{eq:I}
\end{equation}
where the sum goes over both dust species $i$ with the aforementioned mass fractions $f_i$. Here, $B_{\nu}(T)$ is the Planck spectrum (we use the frequency representation) of a blackbody radiator of temperature $T$. The additional factor $\pi a^2$ in the integral over all grain sizes stems from converting the normalised absorption cross section to the absolute value.

For the polarised intensity, we only need to take into account the contribution from silicate grains as we assume that carbonaceous grains are not aligned. If one were to assume perfect alignment for silicate grains starting from a minimal grain size of $a_\textrm{al,min}$ up to a maximum grain size of $a_\textrm{al,max}$, this gives
\begin{equation}
 I_\textrm{pol, 1-phase} = f_\textrm{sil} \times \int_{a_\textrm{al, min}}^{a_\textrm{al, max}} B_{\nu}(T_\textrm{sil}(a))  \times \pi a^2 \times Q_{\textrm{abs, sil} }(a,\lambda) \times a^{-3.5} \textrm{d}a  \, .
 \label{eq:Ipol}
\end{equation}
Note that $a_\textrm{al,max}$ is given by the Larmor limit \citep[see][for more details on the alignment]{Seifried19}. From previous calculations using realistic 3D simulations of molecular clouds \citep[][]{Seifried19}, we found that  $a_\textrm{al,max}$ of the order of the maximum grain size of the dust distribution, i.e. the upper size limit for aligned grains is of less importance, whereas \mbox{$a_\textrm{al, min} \simeq$ 10 nm} (see their figure~13). We note that in reality silicate grains will not be perfectly aligned. This is intrinsically taken into account in the POLARIS toy model (Section~\ref{sec:polarismodelbenchmark}) and will thus reduce the amount of polarised intensity.

\subsubsection{The two-phase case}
\label{sec:2phasetheory}

We can now easily extend the one-phase case by assuming that there is a second phase along the line of sight with different dust temperatures and densities, which contributes to the observed emission. Possible motivations for a two-phase model are e.g. an HII region embedded in a molecular cloud or the gas at the edge vs. the interior of a molecular cloud (see also Section~\ref{sec:motivation}).

In order to properly account for the relative contributions to the emission from both phases, we must first state by which factor the gas and (due to the assumption of a constant dust-to-gas ratio) also the (dust) density is scaled with respect to the first phase. We denote this factor as $f_\textrm{scale}$. We assume that both phases have the same extent along the line of sight\footnote{Assuming a different line-of-sight length can be simply incorporated by adapting $f_\textrm{scale}$ under the assumption that the emission is everywhere optically thin.}. Hence, $f_\textrm{scale}$ is a measure for both the density and column density ratio. In addition, we allow for a further complication by assuming that in this second phase the amount of carbonaceous grains is reduced with respect to the canonical value of \mbox{$f_\textrm{carb}$ = 37.5\%}, which we account for by an additional multiplicative factor, $f_\textrm{scale, carb}$. We emphasise that this latter factor only reduces the amount of carbonaceous grains in the second phase, but leaves the amount of silicate grains unaffected. This approach is motivated by indications that under certain circumstances carbonaceous grains might get destroyed (see Section~\ref{sec:carbondestruction}). Beside the change in density, this second phase can also have different dust temperatures given by $T_{i, \textrm{2}}$.

With this addition, we can now calculate the total and the polarised intensities of the two-phase model as
\begin{flalign}
& I_\textrm{2-phase}  =   I_\textrm{1-phase} \nonumber \\ 
& +  f_\textrm{scale} \times f_\textrm{sil} \times \int_{a_\textrm{min}}^{a_\textrm{max}} B_{\nu}(T_\textrm{sil, 2}(a))  \times \pi a^2 \times Q_{\textrm{abs, sil}}(a, \lambda) \times a^{-3.5} \textrm{d}a \nonumber  \\
&        + f_\textrm{scale} \times f_\textrm{scale, carb} \times f_\textrm{carb} \times \nonumber  \\
&        \int_{a_\textrm{min}}^{a_\textrm{max}} B_{\nu}(T_\textrm{carb, 2}(a))  \times \pi a^2 \times Q_{\textrm{abs, carb}}(a, \lambda) \times a^{-3.5}  
\label{eq:I2phase}
\end{flalign}
and
\begin{flalign}
&I_\textrm{pol, \, 2-phase} =  I_\textrm{pol, \, 1-phase} +  f_\textrm{scale} \times f_\textrm{sil} \times & \nonumber \\ 
& \int_{a_\textrm{al, min, phase 2}}^{a_\textrm{al, max, phase 2}} B_{\nu}(T_\textrm{sil, 2}(a))  \times \pi a^2 \times Q_{\textrm{abs, sil}}(a, \lambda) \times a^{-3.5} \textrm{d}a \, . &
\label{eq:Ipol2phase}
\end{flalign}
Here, $I_\textrm{1-phase}$ and $I_\textrm{pol, \, 1-phase}$ are the intensities originating from the first phase given by Eqs.~\ref{eq:I} and~\ref{eq:Ipol}. In order to keep the number of free parameters at a reasonable level, we use the same values for $a_\textrm{min}$ and $a_\textrm{max}$ as in the first phase. In contrast to that, the minimal and maximum aligned grain sizes $a_\textrm{al, min, phase 2}$ and $a_\textrm{al, max, phase 2}$ can differ from their corresponding values in the first phase, given the different physical conditions. Using POLARIS, these differences between the two phases will automatically be accounted for. If we assume that the emission is optically thin, it is of no relevance whether phase 2 is located in front or behind phase 1 (as seen from the observer). We note that this model could in principle be extended to contain $N$ independent phases along the line of sight as long as the optically-thin assumption holds.

Finally, the wavelength-dependent polarisation degree is given by
\begin{equation}
 p = \frac{I_\textrm{pol}}{I} \, .
 \label{eq:p}
\end{equation}
Note that we have dropped the subscript "1/2-phase" as the formula applies accordingly to both the one- and two-phase model.

\subsection{The POLARIS toy model}
\label{sec:polarismodelbenchmark}

In the following we describe how the toy model described before is implemented in POLARIS \citep{Reissl16}. First, as POLARIS requires an overall density to be set, we use a spatially constant density of 100~cm$^{-3}$ in phase~1 with a mean molecular weight of 2.3 typical for molecular clouds. We use a dust-to-gas ratio of 0.01 and the same dust mass fractions $f_i$ for silicate and carbonaceous grains as stated in Section~\ref{sec:1phasetheory}. We also allow for \textit{individual} dust temperatures of silicate and carbonaceous grains. We emphasise that the latter requires special attention when setting up the POLARIS input files and its script files as otherwise only an averaged dust temperature is calculated (see Appendix~\ref{sec:appendixa} for details).

The grain size distribution is, as stated before, proportional to $a^{-3.5}$, with a lower and upper cut-off radius of \mbox{$a_\textrm{min}$ = 5~nm} and \mbox{$a_\textrm{max}$ = 500~nm}, respectively. We use the dust data provided with the POLARIS distribution named \textsc{silicate\_oblate} and \textsc{graphite\_oblate}, which were created with DDSCAT \citep{Draine13}. We use a 3-dimensional domain, which is cubic in size and has an extent of 15.6~pc, i.e. (15.6~pc)$^3$ in volume, resulting in a total mass of about $21.5 \times 10^{3}$ M$_\odot$. The magnetic field is set to a constant value of 3 $\mu$G and is oriented along the $x$-direction. We use a cubic grid with $2 \times 2 \times 2$ cells in each direction to be able to set up the two phases. In case that two phases are used, each of them spans  $2 \times 1 \times 2$ cells, with the phase change occurring along the $y$-axis. Throughout the paper we will consider observations along the $y$-axis, thus leading to the two-phase case as described in Section~\ref{sec:2phasetheory}.

With POLARIS we have the option to calculate the dust temperatures self-consistently using a Monte-Carlo approach and assuming that the cloud is embedded in an interstellar radiation field (ISRF). However, for the sake of simplicity and in order to have full control over the parameters of the toy model, we manually set the dust temperatures for silicate and carbonaceous grains to size-independent values of $T_\textrm{sil}$ and $T_\textrm{carb}$ as well as $T_\textrm{sil, 2}$ and $T_\textrm{carb, 2}$. These values are constant throughout each phase, but different for silicate and carbonaceous grains.  As stated before, we use POLARIS to calculate $a_\textrm{al, min}$, $a_\textrm{al, max}$, $a_\textrm{al, min, phase 2}$, $a_\textrm{al, max, phase 2}$  for the silicate grains, which then determines the amount of polarised intensity (Eqs.~\ref{eq:Ipol} and~\ref{eq:Ipol2phase}, taking into account imperfect alignment). We assume that silicate grains become aligned with the magnetic field due to radiative torques \citep[][]{Lazarian07,Hoang09}. For the calculation of the radiative torque we use the ISRF of \citet{Mathis83} scaled by a factor of 1.47 such that its energy content corresponds to 1 Draine field \citep{Draine78}. Note that the calculation of $a_\textrm{al, min}$ ($a_\textrm{al, min, phase 2}$) uses the given dust temperature, i.e. POLARIS does not change $T_\textrm{sil}$ ($T_\textrm{sil, 2}$) any more. Carbon grains are assumed to not align with the magnetic field.

\subsection{The 3D MHD cloud model}
\label{sec:3Dmodel}

As a final step in this work, we use a molecular cloud simulated with FLASH \citep{Fryxell00,Dubey08} within the SILCC-Zoom project \citep{Seifried17}. The cloud is described in detail in \citet{Seifried19,Seifried20a} where it is denoted as MC1-MHD. In the following we give only a very brief overview of the simulations, for more details we refer to the aforementioned publications. The cloud forms from the multi-phase ISM modelled within the SILCC project \citep{Walch15,Girichidis16}, where we model a part of a galactic disk with solar neighbourhood properties including supernova feedback, radiative transport and a chemical network to capture the correct thermochemical evolution of the gas. The initial magnetic field strength in the midplane was set to 3~$\mu$G. During the course of the simulation we zoom-in on a forming molecular cloud and resolve it with  a resolution of up to 0.12~pc. At the time the cloud has formed, the magnetic field in the cloud region has an average field strength of about 4.4~$\mu$G. For more details on the simulations we refer to the aforementioned papers. In the following, we briefly describe the POLARIS simulations performed to generate the dust polarisation maps of the cloud MC1-MHD.

We extract a cube with a size of 125~pc centred on the simulated molecular cloud to which we apply the dust polarisation radiative transfer. The dust properties are identical to those mentioned in Section~\ref{sec:1phasetheory} with a dust-to-gas ratio of 0.01 like for the POLARIS toy model. However, in contrast to the toy model, we calculate the dust temperature via POLARIS using a Monte-Carlo approach with a specified radiation field (see below). We calculate size-dependent temperatures for silicate and carbonaceous grains \textit{separately}, which is obtained by setting the command "\textsc{$<$full\_dust\_temp$>$ 1}" in the POLARIS script. We again put special emphasis on the fact that a differentiation between silicate and carbon grain temperatures requires to modify the POLARIS input files before the Monte Carlo simulation, as with the default settings POLARIS will calculate a single (but still size-dependent) temperature for silicate and carbonaceous grains together (see Appendix~\ref{sec:appendixa}). As for the toy model (Section~\ref{sec:polarismodelbenchmark}), we use the dust data provided by POLARIS where carbonaceous grains are assumed to not align.

The simulation does currently not include stellar feedback. In order to still mimic its effect in a simplistic manner, for the radiative post-processing we assume that the cloud is exposed to strong stellar radiation, e.g. from a nearby O/B cluster. Specifically, for the temperature and alignment calculation (again by radiative torques), we apply an ISRF set to 1.47 $\times$ 10$^4$ times the strength of the local ISRF given by \citet{Mathis83} corresponding to 10$^4$ Draine fields as could be found in the vicinity of a cluster of O/B stars. For reference we also use a model with an ISRF of 1 Draine field. In the final radiative transfer step, the emission maps are calculated for wavelengths of 70, 100, 130, 160, 250, 350, 500, 850, 1000, and 1300~$\mu$m to cover the interesting range of the polarisation spectrum, within which the V-shape occurs in some observations (typically with the minimum around $\sim$300 $\mu$m).

\section{Results of the one-phase model}
\label{sec:onephase}

We first discuss the results of our POLARIS toy model (Section~\ref{sec:polarismodelbenchmark}) concentrating on the one-phase case (Section~\ref{sec:1phasetheory}) to demonstrate its basic functionality. As stated before, we consider size-independent dust temperatures, which can, however, differ for silicate and carbon grains.

\subsection{The impact of different silicate  temperatures}
\label{sec:tdustvary}

\begin{figure}
\centering
\includegraphics[width=\linewidth]{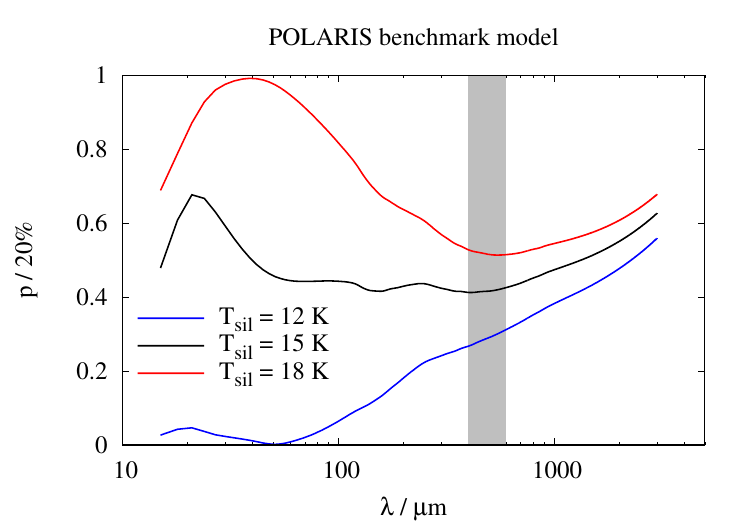} 
\caption{Polarisation spectrum (normalised to a polarisation degree of 20\%) obtained from the POLARIS toy model with one phase only, for three different silicate temperatures ($T_\textrm{carb}$ is kept fix at 15~K). Below \mbox{$\lambda \simeq$ 500 $\mu$m} (indicated by the grey band), the polarisation spectrum strongly depends on the relative difference between the silicate and carbon grain temperature.}
\label{fig:polaris}
\end{figure}

In a first step, we explore the effect of changing $T_\textrm{sil}$ on the polarisation spectrum while $T_\textrm{carb}$ remains fixed at 15~K. The results are shown in Fig.~\ref{fig:polaris}, where we show the cases for $T_\textrm{sil}$ = 12, 15 and 18~K. We renormalise the obtained polarisation degree to 0.2 and consider wavelengths from 15~$\mu$m to 3~mm.

Below \mbox{$\lambda \simeq$ 500 $\mu$m} (indicated by the grey band), the three spectra show very distinct behaviours: For \mbox{$T_\textrm{sil}$ = 12 K}, the polarisation spectrum increases with increasing $\lambda$, whereas for \mbox{$T_\textrm{sil}$ = 18 K} it declines; for \mbox{$T_\textrm{sil}$ = $T_\textrm{carb}$ = 15 K} it remains almost flat down to \mbox{$\lambda \simeq 40 \mu$m}. Above \mbox{$\lambda \simeq$ 500 $\mu$m},  however, the polarisation spectrum starts to increase again for all three cases considered.

These fundamental differences in the polarisation spectrum for short wavelengths also match those discussed by \citet{Hildebrand99} (see also Section~\ref{sec:previouswork} for more details). Furthermore, for the case of $T_\textrm{sil}$ < $T_\textrm{carb}$ our results also show a good qualitative match with those of \citet[][see their models~1 and 3 in their figure~8\footnote{Note that their models~2 and 4 should not be used for comparison as there carbonaceous grains were assumed to contribute to the polarised emission.}]{Draine09}. Furthermore, the qualitative behaviour of the spectrum for $T_\textrm{sil}$ < $T_\textrm{carb}$ showing an increase at \mbox{$\lambda \lesssim$ 500 $\mu$m} also matches the results from \citet{Ashton18} and those from a more complex model discussed by \citet[][see their figure~13]{Bethell07}, although for both authors the spectrum remains rather flat above 500~$\mu$m (see also Section~\ref{sec:modification}).

We note that optical depth effects play only a negligible role in the shape of the spectra. We tested this by rerunning the POLARIS toy model with a 100 times lower density of 1~cm$^{-3}$. The obtained results are almost identical to that of the higher-density runs. Only marginal relative changes of the order of a few 1\% are recognisable, mainly at short wavelengths, i.e. where dust emission is expected to become optically thick more quickly. We also checked a POLARIS toy model where we assumed perfect grain alignment (by setting ``\textsc{$<$ align $>$ alig\_pa}'' in the POLARIS script file). This gives qualitatively identical results (not shown here) as in Fig.~\ref{fig:polaris}; only the polarisation degree is higher by a factor of $\sim$2.5. Hence, the results obtained here can thus be considered as rather robust and intrinsic to the underlying dust temperature distribution ($T_\textrm{sil} \gtrless T_\textrm{carb}$).

These findings have important implications: In the presence of only \textit{one} emitting phase with a given $T_\textrm{sil}$ and $T_\textrm{carb}$, the polarisation spectrum can only have a declining slope, i.e. $p$ dropping with increasing $\lambda$, if silicate grains are \textit{warmer} than carbonaceous grains. This is of particular importance given the fact that there exist observations with a (partly) declining slope of the polarisation spectrum towards a number of sources \citep{Vaillancourt02,Vaillancourt08,Vaillancourt12,Zeng13,Tram24}. However, dust models indicate that silicate grains are generally cooler than carbonaceous grains \citep[e.g][]{Draine85,Guhathakurta89,Li01,Bethell07,Draine11}. We will come back to this peculiar finding later in Section~\ref{sec:twophase}.

\subsection{Influence of the spectral index}
\label{sec:modification}

As mentioned before, some works find a rather flat polarisation spectrum above $\sim$500~$\mu$m \citep{Bethell07,PlanckXXII,Ashton18}, whereas in our case the spectrum is increasing (see Fig.~\ref{fig:polaris}). As we will show in the following, this can be attributed to differences in the dust spectral index $\beta$ used for the absorption cross sections (Eq.~\ref{eq:Qabs}).

\begin{figure}
\centering
\includegraphics[width=\linewidth]{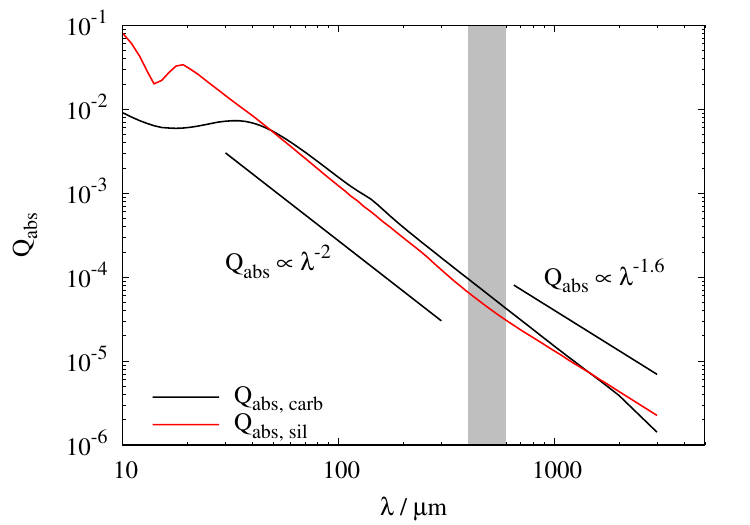} 
\caption{Absorption cross section of silicate and carbonaceous grains with a size of 50~nm used in POLARIS. Around \mbox{$\lambda \simeq$ 500 nm}, indicated by the grey band, the spectral index $\beta_\textrm{sil}$ for silicate grains changes from $\sim$2 to $\sim$1.6.}
\label{fig:Qabs}
\end{figure}
In Fig.~\ref{fig:Qabs} we show the absorption cross section of a silicate and a carbonaceous grain with a size of 50~nm used in our POLARIS model \citep[obtained with DDSCAT\footnote{These are the data contained in the files \textsc{silicate\_oblate.dat} and \textsc{silicate\_oblate.dat} coming with the POLARIS distribution.},][]{Draine13}. As can be seen, for silicate grains the absorption cross section scales roughly as $\lambda^{-2}$ for \mbox{$\lambda$ $<$ 500 $\mu$m}  and as $\lambda^{-1.6}$ above that wavelength. For carbonaceous grains the scaling of $Q_\textrm{abs}$ in POLARIS remains proportional to $\lambda^{-2}$ over the entire wavelength range.

Hence, in order to test the impact of the dust spectral index, we numerically evaluate the one-phase toy model described in Section~\ref{sec:1phasetheory}. For this purpose, we use $Q_\textrm{0, sil} = 1.4 \times 10^{-2}$ and $Q_\textrm{0, carb} = 1 \times 10^{-2}$ in Eq.~\ref{eq:Qabs}. For the spectral index of carbonaceous grains we assume $\beta_\textrm{carb} = 2$ throughout the entire wavelength range. For silicate grains, we use \mbox{$\beta_\textrm{sil}$ = 2} for \mbox{$\lambda \leq$ 500 $\mu$m}. For \mbox{$\lambda \geq$ 500 $\mu$m} we differentiate two cases: In the first case, we set \mbox{$\beta_\textrm{sil}$ = 1.6} above 500~$\mu$m, in agreement with the dust model used in POLARIS (Fig.~\ref{fig:Qabs}). In this case, in addition we replace $Q_\textrm{0, sil}$ in Eq.~\ref{eq:Qabs} by 
\begin{equation}
Q_\textrm{0, sil} = 1.4 \times 10^{-2} \times \left(\frac{500 \, \mu \textrm{m}}{100 \, \mu \textrm{m}}\right)^{-0.4} \; \; \textrm{ if } \; \; \lambda \geq 500 \, \mu\textrm{m} \, ,
\label{eq:Qabssil}
\end{equation}
in order to assure that $Q_\textrm{abs, sil}$ varies continuously across the entire wavelength range. In the second case, we keep \mbox{$\beta_\textrm{sil}$ = 2} above 500~$\mu$m (and thus also $Q_\textrm{0, sil}$ remains unchanged).

\begin{figure}
\centering
\includegraphics[width=\linewidth]{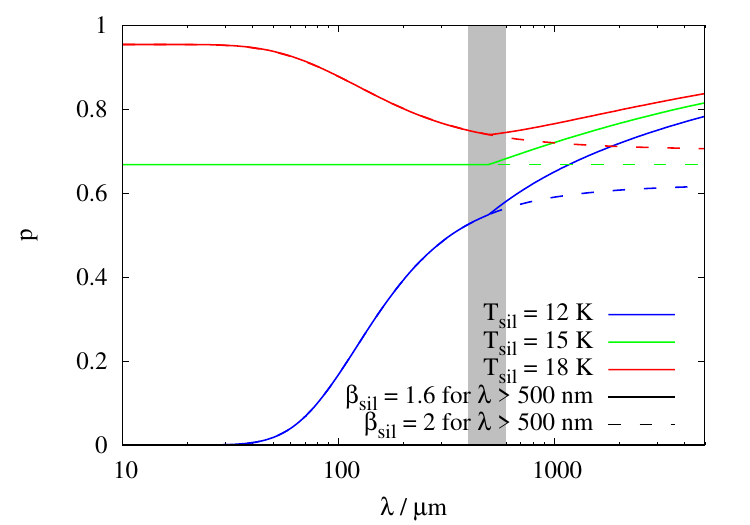} 
\caption{Polarisation spectrum of our numerically evaluated one-phase model for different silicate dust temperatures ($T_\textrm{carb}$ is kept fixed at 15~K) and two choices for the spectral index $\beta_\textrm{sil}$ (see text). Changing the spectral index $\beta_\textrm{sil}$ from 2 to 1.6 above \mbox{$\lambda$ = 500 $\mu$m} motivated by the dust data provided by POLARIS (see Fig.~\ref{fig:Qabs}) leads to the increase in this wavelength range, in good agreement with the POLARIS results (compare Fig.~\ref{fig:polaris}). In contrast, for an unchanged $\beta_\textrm{sil}$ of 2 (dashed lines), the spectrum would be almost flat.}
\label{fig:onephase_spectralindex}
\end{figure}
We plot the results of our one-phase model by numerically integrating Eqs.~\ref{eq:I} and~\ref{eq:Ipol} in Fig.~\ref{fig:onephase_spectralindex}, again using a carbon dust temperature of \mbox{$T_\textrm{carb}$ = 15 K} and three values for the silicate dust temperature of \mbox{$T_\textrm{sil}$ =} 12, 15, and 18~K, respectively. The dust temperatures thus match those used in the POLARIS toy models. Considering first the case, where $\beta_\textrm{sil}$ changes above \mbox{$\lambda =$ 500 $\mu$m}, we find a reasonable qualitative match of the polarisation spectra with those of the POLARIS models (see Fig.~\ref{fig:polaris}). In particular, above \mbox{$\lambda$ $\simeq$ 500 $\mu$m} (indicated by the grey band) the polarisation spectrum shows a clear change in its slope: The polarisation degree increases by up to a few 10\% relative to the value of $p$ at 500 $\mu$m, although the increase is somewhat less pronounced than for the POLARIS models.

Contrary to that, when assuming that $\beta_\textrm{sil}$ remains at a value of 2 above \mbox{$\lambda$ = 500~$\mu$m} (dashed lines in Fig.~\ref{fig:onephase_spectralindex}), the spectrum is almost flat in that wavelength range. This matches well with the results of \citet{Ashton18}, who use a constant $\beta_\textrm{sil}$ over the entire wavelength range\footnote{We note that \citet{Ashton18} use \mbox{$\beta_\textrm{sil} =1.6$} which seems, however, to not match with the dust data given by \citet{Draine13} and that used in POLARIS. Using $\beta_\textrm{sil}$ = 1.6 for all $\lambda$ in our numerical model would, however, also result in a flat spectrum at long wavelengths.}.

These results demonstrate the importance of the spectral index for the shape of the polarisation spectrum. In order to obtain an increase in the polarisation spectrum at wavelengths above $\sim$500~$\mu$m, a change in the spectral index of silicate grains around this wavelength is required, while the spectral index of carbonaceous grains has to remain the same\footnote{We checked that, when also changing $\beta_\textrm{carb}$ to 1.6 above $\lambda \simeq$ 500~$\mu$m, we obtain the same results as if both $\beta$ remained fixed at a value of 2.}.

\section{The two-phase model}
\label{sec:twophase}

Investigating the V-shaped polarisation spectra found in various observations \citep[][and reference therein]{Hildebrand00,Vaillancourt07SPIE,Vaillancourt08,Vaillancourt12} reveals a significantly more pronounced V-shape than that shown in Fig.~\ref{fig:polaris}. Moreover, for a one-phase model the only way to obtain a declining polarisation spectrum at short wavelengths is to have silicate grains which are \textit{warmer} than carbonaceous grains.  However, as noted before, detailed dust models \citep[e.g.][]{Draine85,Guhathakurta89,Li01,Bethell07,Draine11} indicate that silicate grains should be on average colder than carbonaceous grains. Hence, our tentative requirement of silicate grains being warmer than carbonaceous grains indicates that a one-phase model is not sufficient to explain observed polarisation spectra. For these reasons, we extend our POLARIS one-phase model to a two-phase model as described in Section~\ref{sec:2phasetheory}. As stated before, such a two-phase model also been investigated before by \citet{Hildebrand99} and \citet{Ashton18}. However, here we try to explore in a more comprehensive way the effect of various parameters, which can influence the polarisation spectrum.

\subsection{Physical conditions in the two phases}
\label{sec:motivation}

The usage of a model with two different (dust temperature) phases can be motivated by two scenarios:
\begin{enumerate}
\item First, consider a cold molecular cloud (phase 1) which is illuminated from the outside by a nearby massive star. The stellar radiation will heat up the outer and less dense regions of the molecular cloud (phase 2), whereas the central parts of the cloud remain relatively well shielded from the radiation.
\item The second scenario comprises a HII region, whose density is lowered due to its ongoing expansion (phase 2) and which is embedded in a dense molecular cloud (phase 1).
\end{enumerate}
Hence, in both cases we expect a more dilute phase~2 with dust grains being warmer than in the colder and denser phase~1. As our polarisation spectrum is not affected by optical depth effects (Section~\ref{sec:tdustvary}), we can treat both cases equivalently with our two-phase model. We note again that the length along the line of sight for both phases is equal and set to half the domain size, i.e. 7.8~pc.

The motivation for the choice of these two phases lies in the fact that molecular cloud regions like OMC-1, W51, and NGC2024, which show a V-shaped polarisation spectrum \citep{Vaillancourt07SPIE,Vaillancourt08,Vaillancourt12,Tram24}, are indeed associated with nearby massive star formation. We thus expect physical conditions reminiscent of a dense and cold phase in the center of the cloud complex and a warm and dilute phase in its surroundings. 

The dilution in the warm (i.e. the second) phase is given in our toy model by the factor $f_\textrm{scale}$ (compare Eqs.~\ref{eq:I2phase} and~\ref{eq:Ipol2phase}). In the following we consider a range of dilution factors 
\begin{equation}
f_\textrm{scale} = 10^{-1} - 10^{-4} \, ,
\end{equation} 
with a fiducial value of 10$^{-3}$. Hence, in the POLARIS toy model the second phase has a density of \mbox{$f_\textrm{scale} \times$ 100 cm$^{-3}$}.

In addition, we assume that the grains are heated up by a constant factor, $f_\textrm{heat}$\footnote{This constant factor is chosen in order to keep the number of free parameters at a reasonable level. Using grain temperature values other than those scaled up by a constant factor does, however, only quantitatively, but not qualitatively influence the results.}, compared to the cold phase~1, for which we use fiducial values of \mbox{$T_\textrm{sil} = 12$ K} and \mbox{$T_\textrm{carb}$ = 15 K}. I.e., in both phases we always keep \mbox{$T_\textrm{sil} < T_\textrm{carb}$}. We use heat-up factors of \mbox{$f_\textrm{heat}$ = 2 -- 5}, resulting in dust temperatures in phase~2 of
\begin{eqnarray}
T_\textrm{sil, 2} &=& 24 - 60 \, \textrm{K \;\; and} \nonumber \\
T_\textrm{carb, 2} &=& 30 - 75 \, \textrm{K} \nonumber  \, ,
\end{eqnarray}
with a fiducial value of \mbox{$f_\textrm{heat}$ = 3}. Such values can be reached in the presence of nearby O/B stars which raise the ISRF by a factor of 100 -- 1000 compared to a standard Draine field \citep[e.g.][]{Wolfire94,Draine11}. The values also cover the range of dust temperatures found in HII regions \citep[e.g.][]{Anderson12,Relano16}. We note that also self-consistently calculated dust temperatures of our POLARIS toy model (Section~\ref{sec:polarismodelbenchmark}), which were obtained with a Monte-Carlo approach using an elevated ISRF with a strength of 100 and 1000 times a standard Draine field, lie in this range. The additional scaling factor for carbonaceous grains is kept for the moment at \mbox{$f_\textrm{scale, carb}$ = 1}.

\subsection{Results of the two-phase model}
\label{sec:twophaseresult}

\begin{figure*}
\centering
\includegraphics[width=0.9\linewidth]{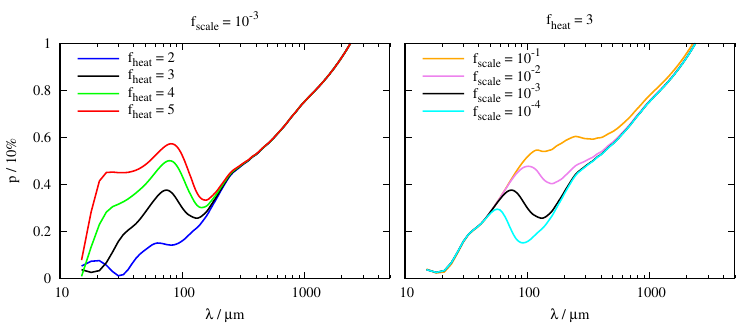} 
\caption{Polarisation spectra obtained with the POLARIS two-phase model varying the dilution factor ($f_\textrm{scale}$, left panel) and the heating-up of dust ($f_\textrm{heat}$, right panel) in phase~2. In both cases, a flattening and a minimum around $\lambda$ = 100 - 200 $\mu$m can be found, similar to observed V-shaped polarisation spectra. For larger differences between both phases, i.e. for a warmer and more dilute phase~2, the minimum is more pronounced.}
\label{fig:twophase_test}
\end{figure*}
We show the results of the two-phase toy model obtained with POLARIS with the afore described parameters in Fig.~\ref{fig:twophase_test}. The shape of the polarisation spectrum deviates clearly from the behaviour found in the one-phase models with \mbox{$T_\textrm{sil} < T_\textrm{carb}$} (see Fig.~\ref{fig:polaris}). In particular, the two-phase models exhibit a more or less pronounced minimum of the spectrum around $\lambda$ = 100 - 200 $\mu$m (except for the case with \mbox{$f_\textrm{scale} =10^{-1}$} and $f_\textrm{heat} = 3$). In general, we find that the stronger the contrast between the two phases, i.e. the smaller $f_\textrm{scale}$ and/or the larger $f_\textrm{heat}$, the more pronounced is the dip. This leads to the aforementioned observed V-shape structure in the most extreme cases and thus, to a pronounced deviation from the one-phase model. Furthermore, the relative increase in $p$ at long wavelengths appears to be more pronounced. In addition, we observe a shift of the minimum towards longer wavelengths as $f_\textrm{scale}$ or $f_\textrm{heat}$ increase.

Despite this qualitative match between our simple two-phase model and the observations showing a V-shape polarisation spectrum \citep{Hildebrand00,Vaillancourt07SPIE,Vaillancourt08,Vaillancourt12,Tram24}, the minimum in our polarisation spectrum around $\lambda$ = 100 - 200 $\mu$m appears to be (i) less pronounced and (ii) at somewhat shorter wavelengths than in these observations. In the observational spectra, the minimum has a value of $p$ which is a factor of 1.5 -- 2 below the values at neighbouring wavelengths and appears around \mbox{$\lambda \simeq$ 350 $\mu$m} \citep[see e.g. figure 5 in][for a nice compilation]{Vaillancourt12}. This indicates that our two-phase model might still be missing some important dust properties, which are relevant in the observed regions. We will investigate this possibility in the following two sections. The effect of a more realistic density distribution is further investigated in Section~\ref{sec:3Dresult}.

\subsection{Carbon grain destruction in illuminated regions?}
\label{sec:carbondestruction}

Recent results indicate that carbonaceous grains seem to be more rapidly re-processed in the ISM than silicate grains \citep{Jones11,Jones14}. Specifically, very small carbon grains and polycyclic, aromatic hydrocarbons are believed to be destroyed in the presence of strong UV radiation fields \citep[e.g.][]{Rapacioli06,Pilleri12,Pavlyuchenkov13,Egorov23}. Moreover, measuring the gas phase abundance of carbon, some observations find elevated carbon levels in HII regions compared to the average ISM abundance, whereas the silicate abundance in those HII regions seems to remain unchanged \citep[][see also \citealt{Mathis05} for a short overview]{Rubin93,Esteban99}.

Taken together, these results indicate that in the presence of a strong UV radiation field, carbonaceous grains in the ISM might get partially destroyed whereas silicate grains seem to remain (mostly) intact. In our two-phase toy model used in POLARIS, we can emulate this effect by setting the carbon scaling factor $f_\textrm{scale, carb}$ to a value below 1, which reduces the amount of carbonaceous grains in the dilute and warm, i.e. in the illuminated phase~2, while the amount of silicate grains remains unchanged.

\begin{figure*}
\centering
\includegraphics[width=0.9\linewidth]{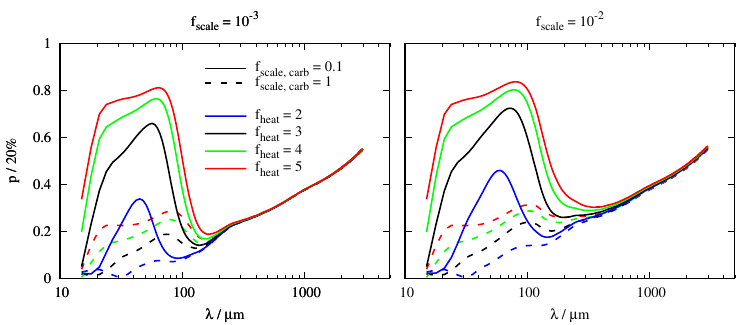} 
\caption{Polarisation spectra for the two-phase model with a dilution factor of $f_\textrm{scale}$ = 10$^{-3}$ (left) and 10$^{-2}$ (right) and various heat-up factors (colour coding). The solid lines show the spectra obtained under the assumption that in the warm, dilute phase~2, the carbon grain content is reduced to 10\% of its fiducial value due to the impinging  UV radiation. This leads to a significantly more pronounced V-shape than for the fiducial case, where carbon grains remain intact (dashed lines).}
\label{fig:twophase_fgrain}
\end{figure*}
In Fig.~\ref{fig:twophase_fgrain} we compare the results obtained for different dilution and heat-up factors when setting $f_\textrm{scale, carb}$ to either 1 (assuming no destruction of carbon grains, dashed lines), or 0.1 (assuming that 90\% of all carbonaceous grains are destroyed, solid lines). It can be clearly seen that in all considered cases the destruction of carbonaceous grains would lead to a more pronounced V-shape in the polarisation spectrum. This is mainly due to the increase of $p$ at short wavelengths. The polarisation degree reached at \mbox{$\lesssim$ 100 $\mu$m} is higher by up to a factor of $\sim$2 compared to the value at the minimum. Also the position of the minimum is shifted slightly towards longer wavelengths compared to the \mbox{$f_\textrm{scale, carb} = 1$} case, occurring at wavelengths of up to 300~$\mu$m for \mbox{$f_\textrm{scale} = 10^{-2}$}, thus improving the match with observations that show the minimum around 350~$\mu$m.

We emphasise that when assuming a perfect alignment of all silicate grains in POLARIS, we obtain results (not shown) which are -- beside a polarisation degree which is scaled up by a factor of $\sim$2.5 over the entire wavelength range -- qualitatively and quantitatively almost identical to those shown so far. Hence, the alignment efficiency seems to play only a minor role for the spectral shape.

\begin{figure*}
\centering
\includegraphics[width=0.85\linewidth]{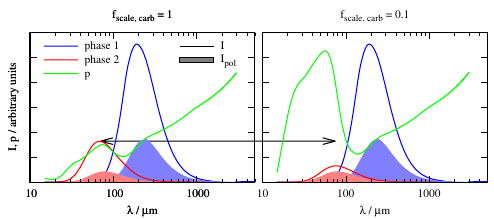} 
\caption{Total (lines) and polarised intensities (shaded areas) in arbitrary units originating from the two phases (blue and red) of one of our two-phase models with \mbox{$f_\textrm{scale}$ = 10$^{-3}$} and $f_\textrm{heat}$ = 3. We show the case where carbon grains in phase~2 remain intact (left) and where 90\% of the carbon grains in phase~2 are destroyed (right). The destruction of carbon grains reduces the amount of total intensity at short wavelengths  originating from phase~2 (indicated by the black arrow to guide the readers eye), leading to a larger polarisation degree $p$ in this range (green curve).}
\label{fig:origin}
\end{figure*}
The origin of the increase in $p$ at short wavelengths when reducing the carbon grain content in phase~2, can be understood by considering the total and polarised intensity in Fig.~\ref{fig:origin}. The assumed destruction of carbon grains (right panel) leads to a reduction of the total intensity emitted from phase~2, which dominates the emission at wavelengths below $\sim$100~$\mu$m. As all other quantities remain unchanged, this in turn results in an increased polarisation fraction in this range.

As noted before, in observations showing a pronounced V-shape in the polarisation spectrum, there is massive star formation going on in the vicinity of the observed molecular clouds (OMC1, W51, NGC2024), possibly leading to an elevated radiation field in their environment. Given that our polarisation spectra tend to match these observations better, if we assume a reduced carbon grain content in the dilute and warm phase~2, this indicates that carbon grain destruction might indeed be at work in regions illuminated by strong radiation fields, as indicated by the observations mentioned in the beginning of this section.

\subsection{A 3D multi-phase cloud model}
\label{sec:3Dresult}

In real molecular clouds with star formation and feedback, the distribution of the gas in the temperature-density phase space is significantly more complex than in our two-phase mode, spanning several orders of magnitude in both temperature and density \citep[see e.g.][for the phase diagrams of the SILCC-Zoom simulations]{Walch15,Seifried20b}. For this reason, we explore the polarisation spectrum returned by POLARIS when applied to a 3D, MHD simulation of a molecular cloud (see Section~\ref{sec:3Dmodel}). This will improve the geometrical simplifications inherent to the 1D toy model discussed so far. However, the 3D, MHD model still contains simplifications concerning the star formation process. For this reason, this section rather serves as a proof of concept and a more in-depth analysis will be deferred to a subsequent paper.

\begin{figure}
\centering
\includegraphics[width=\linewidth]{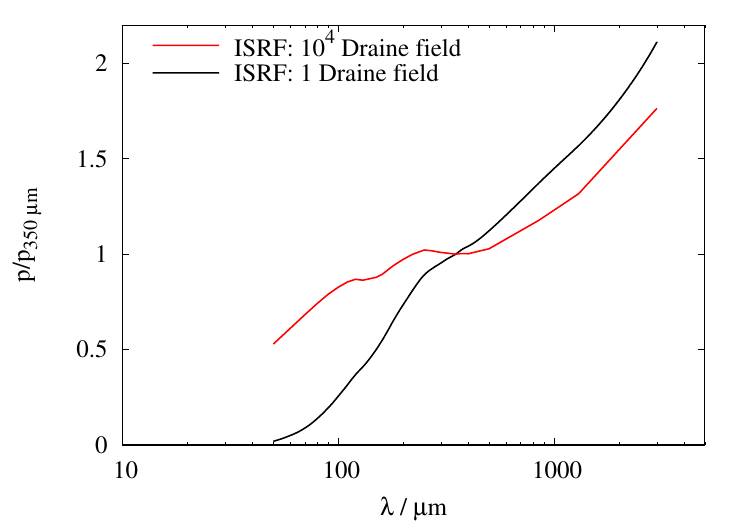} 
\caption{Polarisation spectra calculated with POLARIS using a 3D, MHD molecular cloud simulation. The two lines are obtained for different radiation field strength used for the calculation of the dust temperature and the grain alignment. Overall, for a strongly elevated radiation field we obtain a potential weak minimum in the polarisation spectrum around \mbox{$\lambda \simeq 350$ $\mu$m} which is, however, significantly less pronounced than in the observations shown in Fig.~\ref{fig:obs}.}
\label{fig:ISRF}
\end{figure}

For demonstrative purposes, the complex column density structure of the cloud and the obtained polarisation degree at 100 and 350~$\mu$m are shown in Fig.~\ref{fig:map}.
In Fig.~\ref{fig:ISRF}, we show the resulting polarisation spectra for the case that the cloud is irradiated by (i) a high ISRF  with $G_0 = 1.7 \times 10^4$ (red line) and (ii) a moderate ISRF typical for the solar neighbourhood with $G_0 = 1.7$ (black line). For each wavelength, we first calculate the ratio of $p/p_\textrm{350 $\mu$m}$ in each pixel and then take the median of the ratio over the entire map, a common approach used in the literature \citep[e.g.][]{Vaillancourt08,Vaillancourt12,Gandilo16,Shariff19}. The spectrum for the standard ISRF shows the expected steady increase as already seen in the POLARIS one-phase toy model (see blue line in Fig.~\ref{fig:polaris}). In contrast to that, the model with an elevated ISRF shows a clear flattening with a potential, weakly pronounced minimum around 350~$\mu$m. The position of the minimum seems to match better with the already discussed observational results than that of our two-phase models presented before (e.g. Fig.~\ref{fig:twophase_test}). The depth of the potential V-shape is, however, significantly lower than that found in the toy models as well as in observations (see Section~\ref{sec:obs-comparison}).

\begin{figure}
\centering
\includegraphics[width=\linewidth]{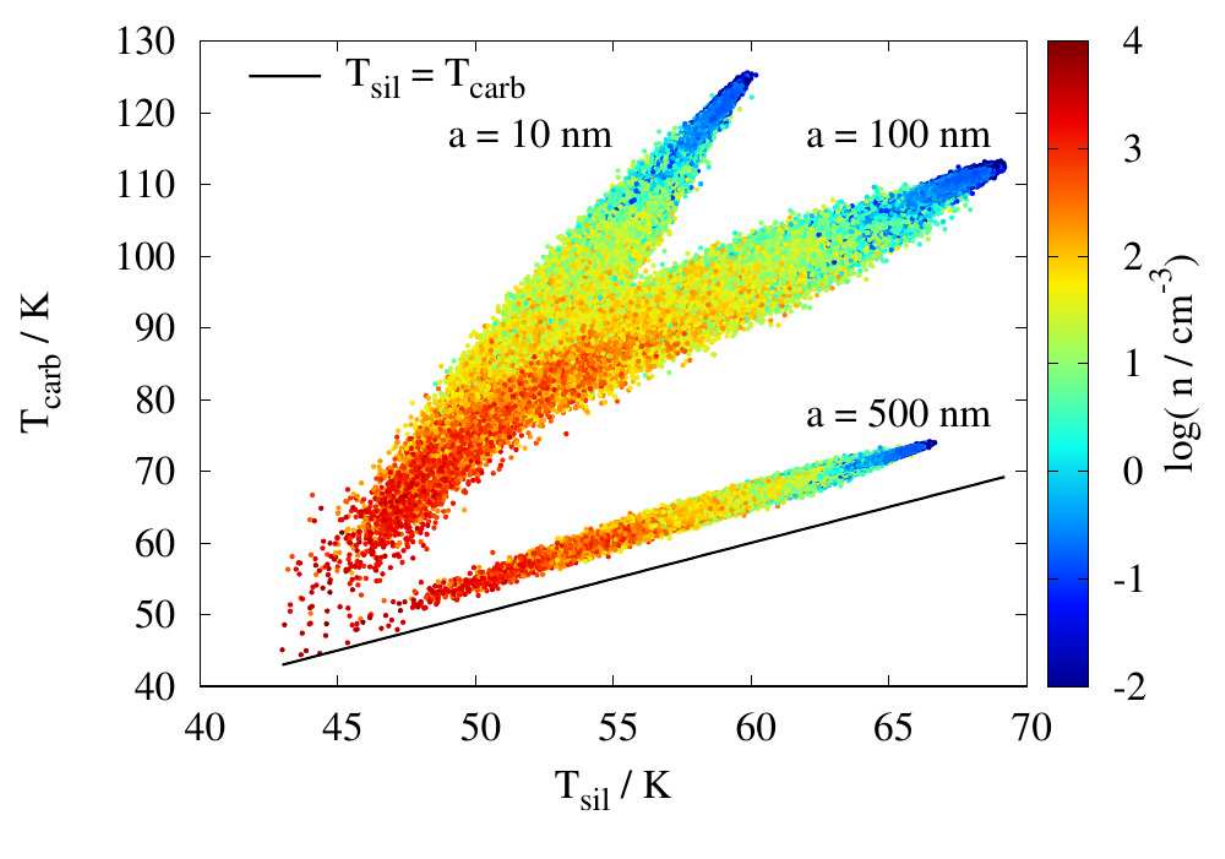} 
\caption{Carbon vs. silicate grain temperatures obtained from the 3D, MHD cloud simulation assuming an ISRF of 10$^4$ Draine fields for grain sizes of 10, 100, and 500~nm. The points are colour-coded with the local density. The dust temperature decreases with increasing density where carbonaceous grains are warmer than silicate grains. To guide the readers eye, we show the line where both temperatures would be identical (black line).}
\label{fig:tdust_polaris}
\end{figure}
In order to understand the origin of the flattening (and the weakly pronounced V-shape) of the polarisation spectrum around 350~$\mu$m for the case of an ISRF of 10$^4$ Draine fields, we analyse in Fig.~\ref{fig:tdust_polaris} the dust temperatures in the simulation cube for both silicate and carbonaceous grains as calculated from POLARIS. The points are colour-coded with the local density. We show the results for grain sizes of 10, 100, and 500~nm. The temperatures are spread over a wide range of $\sim$40~K to 125~K. Carbon grains at a given point are in general warmer than silicate grains, in agreement with other findings \citep{Draine85,Guhathakurta89,Li01,Bethell07,Draine11}, and show a larger spread in temperature. Furthermore, we find decreasing $T_\textrm{d}$ with increasing density for both carbonaceous and silicate grains. In general, the dust temperature distribution thus show a relatively wide range. The relative differences are of a factor of $\sim$1.5 -- 2.5, depending on the considered size and type of the grain. As shown before, such temperature differences are the prerequisite for a V-shaped polarisation spectrum. 

Overall, the fact that the minimum, if present at all, in the synthetic polarisation spectrum obtained from the 3D MHD simulation is significantly lower than those in real observations once more indicates that carbon grain destruction indeed might be required to obtain a V-shape -- something which we plan to explore in more detail in subsequent work. Furthermore, the results indicate that taking into account a more realistic spatial distribution of gas and dust seems to affect the position of the minimum of the polarisation spectrum, shifting it towards longer wavelengths around 350 $\mu$m in the 3D, MHD model, which is thus in better agreement with the observational results.

Finally, we again note that the analysis in this section using a 3D, molecular cloud simulation mainly serves as a proof of concept, both due to the simplified manner with which we mimic the enhancement of the IRSF and the lack of potential carbon grain destruction. Despite this, the observed trend towards a V-shaped polarisation spectrum when enhancing the radiation field, is highly promising as it confirms the postulation obtained from the two-phase toy model (Section~\ref{sec:twophaseresult}). In a follow-up work we therefore plan to investigate this trend more thoroughly using (i) MHD simulations including self-consistent stellar feedback and thus (ii) more sophisticated sources for the radiation field, like e.g. stellar clusters and (iii) possibly also a description for the destruction of carbon grains in POLARIS, as postulated in Section~\ref{sec:carbondestruction}. Furthermore, we plan to also investigate local variations of the polarisation spectrum occurring \textit{within} a single molecular cloud \citep[see e.g. the observations of][]{Santos19,Cox25}.

\section{Discussion}
\label{sec:discussion}

\subsection{Comparison to observations}
\label{sec:obs-comparison}

\begin{figure}
\centering
\includegraphics[width=\linewidth]{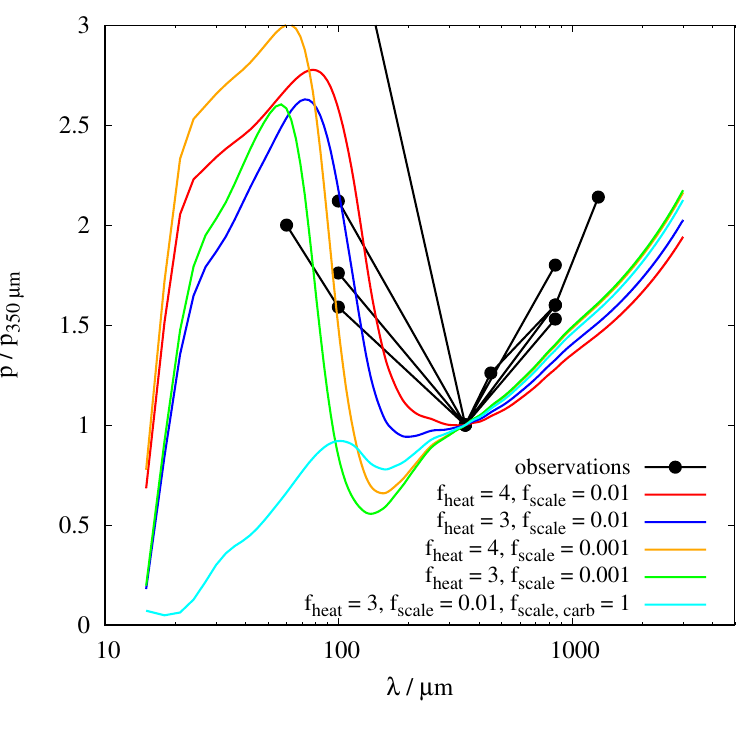} 
\caption{Comparison of observational data (black points) taken from \citet{Vaillancourt07,Vaillancourt12} with results of our two-phase model with selected parameter combinations (here always \mbox{$f_\textrm{scale, carb}$ = 0.1}, except for the cyan line). Overall, our model shows an improved match at short wavelengths if we assume that carbon grains are partially destroyed in the warm, phase~2.}
\label{fig:obs}
\end{figure}

In Fig.~\ref{fig:obs} we show the polarisation spectrum of our two-phase model obtained with POLARIS for a few selected parameter combinations in comparison to observations of W51, M17, NGC2024, DR21, OMC-1 and OMC-3. The data are taken from \citet[][their figure~2]{Vaillancourt07SPIE} and \citet[][their figure~5]{Vaillancourt12}. These regions correspond to molecular clouds with nearby or embedded ongoing massive star formation, which is why we consider the two-phase model applied here as a reasonable approximation to the actual conditions in these regions. The polarisation degree shown in Fig.~\ref{fig:obs} is in all cases normalised to its value at \mbox{$\lambda$ = 350 $\mu$m}. 

There are a few main results to be taken from this comparison. First, in order to roughly match the observed data at low $\lambda$, we require a partial destruction of carbonaceous grains in the warm phase~2 (compare the cyan line with $f_\textrm{scale, carb}$ = 1 to all other lines, where $f_\textrm{scale, carb}$ = 0.1, see also Section~\ref{sec:carbondestruction}). 

Second, we see a tendency that for a density contrast ($f_\textrm{scale}$) between both phases of 0.01, the match is somewhat better than for stronger contrasts ($f_\textrm{scale}$ = 0.001). This might reflect the fact that in reality the density transition from the exterior to the interior of a molecular cloud is rather smooth. Hence, the ``average'' density contrast between the cold interior (average densities around 100~cm$^{-3}$) and warm exterior (diffuse ISM with densities around 1~cm$^{-3}$) cm of the cloud is significantly smaller than the maximum contrast. Moreover, a density contrast of 0.01 would also match the density ratio of established HII regions which are a few 1~Myr old and located inside a molecular cloud with a typical gas density of a few 100~cm$^{-3}$ \citep[e.g.][]{Bisbas15}.

Lastly, our results indicate that additional measurement between 100 to 350~$\mu$m are essential to further constrain the dust properties, as our models show some significant differences in this range. Specifically, with the observational data shown, a minimum at wavelengths lower than 350~$\mu$m cannot be excluded. We note, however, that recent observations of OMC~1 by \citet{Tram24} indicate that indeed the minimum is reached at wavelengths of 200~$\mu$m or above (see their figure~6), thus again favouring our models with a smaller (column) density contrast ($f_\textrm{scale}$ = 0.01).

In the long-wavelength range our toy models tend to slightly underpredict the (normalised) polarisation degree (Fig.~\ref{fig:obs}). The slope appears to be somewhat shallower than that of the observational data and rather independent of the chosen model parameters. Such an underprediction in the slope was also found e.g. in \citet[][see their figure~3]{Vaillancourt08}. Possible reason for this are manifold. Firstly, we still deal with a highly simplified 1D geometry (see also Section~\ref{sec:3Dresult}). This applies both to the density as well as temperature distribution of the dust, which are oversimplifications of the complex environment of star forming regions in molecular clouds. We thus try to make a first step towards more realistic conditions in Section~\ref{sec:3Dresult}. Also the dust properties might not be captured in their entire detail which might affect the shape of the polarisation spectrum.  As discussed in Section~\ref{sec:modification}, the spectral index can have a significant effect on the steepness of the spectral slope at long wavelengths. For example \citet{Dupac03} report an anticorrelation between the dust temperature and the spectral index that we do not account for here.

In summary, the qualitative comparison of our two-phase model with actual observations showing a V-shaped polarisation spectrum results in the following key points:
\begin{enumerate}
\item A two-phase medium with a cold, dense and a warm, dilute phase is always needed to explain the V-shape.
\item The depth of the observed V-shape indicates that carbon grains might have been destroyed by an elevated radiation field.
\end{enumerate}

\subsection{Uncertainties in the spectrum at long wavelengths}

Observations of the V-shaped spectrum are sparse \citep{Hildebrand00,Vaillancourt07SPIE,Vaillancourt08,Vaillancourt12,Lopez22,Tram24}. In addition, due to instrumental constraints, these spectra are often obtained by combining results from different observational facilities in order to cover the desired large wavelength range. As pointed out by \citet{Cox25}, this could introduce systemic biases in the polarisation degrees obtained, e.g. due to different methods in spatial referencing used to correct for polarised emission from the diffuse fore-/background \citep[see also figure~4 in][]{Fissel16}.

The potential influence of this effect on the polarisation spectrum can be seen in figure~3 of \citet{Tram24}, who observed a V-shaped spectrum in OMC~1. Their datapoints taken with JCMT/POL-2 at 850~$\mu$m differ substantially from those taken at a comparable wavelength of 870~$\mu$m  taken with APEX/PolKA. Specifically, for 4 out of 6 cases,  the latter are substantially lower.  Hence, replacing the result at 850~$\mu$m obtained with JCMT/POL-2 with APEX/PolKA at 870~$\mu$m would result in no significant V-shaped spectrum, which, as pointed out by \citet{Tram24}, emphasises the need of additional observations in this wavelength range. We also note that somewhat lower values of $p$ at long wavelengths would also reduce the apparent underprediction of the slope by our toy model in this range (Fig.~\ref{fig:obs}).

Moreover, the rise at long wavelengths in V-shaped spectra is seemingly in disagreement with (almost) flat spectra at $\lambda >$ 500~$\mu$m obtained e.g. with the Planck satellite \citep[][but see also \citet{Ashton18}]{PlanckXXII}. One obvious reason for this could be that different regions can have different dust properties, like the temperature or size distribution or composition. 
Related to this, a change in the spectral index $\beta$, as discussed in Section~\ref{sec:modification}, can have a significant influence on the long-wavelength range of the spectrum, particularly the difference between that of carbonaceous and silicate grains, $\beta_\textrm{carb} - \beta_\textrm{sil}$. However, this difference is not known \citep{PlanckXXII} and could vary from region to region, e.g. depending on the presence or absence of nearby massive star formation. Furthermore, also the relative temperature differences between carbonaceous and silicate grains can alter the spectrum at long wavelengths, with smaller differences leading to a flatter spectrum (see Fig.~\ref{fig:polaris}). All in all, we speculate that -- beside the observational uncertainties mentioned before -- these differences in dust properties can lead to polarisation spectra varying significantly from region to region.

\subsection{Relation to previous works}
\label{sec:previouswork}

Models suggested so far in the literature have required different alignment properties for different dust \textit{components}\footnote{Note that we here intentionally refer to components rather than phases, see further down for the explanation.}, specifically a correlation between the alignment efficiency and the temperature of each component \citep{Hildebrand99,Hildebrand00}, to explain a rise or drop of the polarisation spectrum. Also other authors emphasise this correlation \citep[e.g.][]{Vaillancourt02,Vaillancourt08,Vaillancourt12}. For example, \citet{Vaillancourt08} assume that in their cold phase the alignment efficiency is zero. On first view this seems to be contradictory to our results, as we do not impose any specific requirements on the alignment properties in both phases, but allow them to adapt to the according physical conditions using POLARIS.

A closer look, however, reveals that this difference could partly be of semantic, rather than of physical origin: For example, \citet{Hildebrand99,Hildebrand00} do not intrinsically differentiate between silicate and carbonaceous grains. The authors assume a good alignment for their component A (i.e. corresponding to our silicate grains) and a small/non-existing alignment for their component B (i.e. corresponding to our carbonaceous grains).  Hence, the \textit{two-component} model presented by \citet{Hildebrand99} factually corresponds to our \textit{one-phase} model. 
Furthermore, in their terminology our two-phase model is actually a 2 (dust type) $\times$ 2 (phase) = 4-component model, where two of the four dust components (here, the warm and cold carbonaceous grains) do not contribute to the polarised emission, i.e. are unaligned.

Beside using for the first time a self-consistent dust polarisation radiative transfer code to calculate polarisation spectra, our considerations thus present a step forward in our intuitive understanding of V-shaped polarisation spectra: In order to explain this V-shape, \citet{Hildebrand00} require a three-component model with specifically chosen temperatures and alignment properties for each component (see their figure~18). However, as we have shown, when accounting for a natural mix of silicate and carbonaceous grains, such a fine-tuning is not required any more. Rather, a two-phase model (or a smooth transition from a dense, cold to a dilute, warm regime) -- motivated by either an externally illuminated or internally heated molecular cloud -- will naturally lead to a V-shaped polarisation spectrum with no specifically tuned requirements for the alignment properties of the grains. Furthermore, to our knowledge our work also presents the first systematic exploration of the effect of environmental conditions on the shape of the polarisation spectrum.

Finally, in \citet{Seifried19} we have shown that even inside cold and dense molecular clouds, also small grains (\mbox{$\sim$10 nm}) might stay well aligned up to densities of $\sim$1000~cm$^{-3}$ (corresponding to a visual extinction of \mbox{$A_\textrm{V} \simeq 3$}), i.e. also in those regions which are reasonably well shielded against the ISRF. Hence, the anti-correlation between dust temperature and alignment efficiency, as required by the aforementioned authors, seems to be in general not very strongly pronounced and thus, would only play a minor role in shaping the polarisation spectrum. This conclusion is in agreement with our findings that, when assuming a perfect alignment of the silicate grains, the polarisation spectrum does not change qualitatively (Section~\ref{sec:carbondestruction}).

\section{Conclusions}
\label{sec:conclusion}

In recent years, dust polarisation observations have reported a particular dependence of the polarisation degree on the wavelength for selected molecular clouds. In these observations, this so-called polarisation spectrum shows a pronounced V-shape, i.e. exhibits a minimum around 350~$\mu$m \citep[][and references therein]{Hildebrand00,Vaillancourt07SPIE,Vaillancourt08,Vaillancourt12,Tram24}. In this work we extend previous theoretical works to gain a better quantitative understanding of the origin of this peculiar shape. 

For this purpose, in a first step we introduce a simple toy model for a single dust phase, that is dust consisting of silicate grains aligned with the magnetic field \textit{and} unaligned carbonaceous grains, both having different temperatures from each other. The toy model is evaluated with the dust polarisation radiative transfer code POLARIS. In this context, we provide a guideline how to incorporate separate dust temperatures for silicate and carbon grains in POLARIS, which requires special consideration. We show that with only one phase, the V-shape found in various observed polarisation spectra cannot be explained.

For this reason, we extend our model to a two-phase model, where we assume that in the second phase the dust grains are warmer and the density is reduced. This is motivated by either molecular clouds illuminated by nearby massive stars or HII regions embedded inside them. In the following we summarise our main results:
\begin{itemize}
\item A two-phase medium, i.e. regions with different dust temperatures and densities along the line of sight, is required to produce a well-pronounced V-shape in the polarisation spectrum. This is in agreement with previous results. However, contrary to these previous works, our model does not require any coupling between the alignment efficiency and the temperature of the phase.
\item The V-shape in the polarisation spectrum is more pronounced when the density and temperature contrast between the cold, dense phase and the warm, dilute phase is larger.
\item If carbon grains are destroyed in the second (warm and dilute) phase by UV radiation from nearby massive stars, the characteristic of the V-shape is significantly enhanced, i.e. the polarisation spectrum shows a pronounced minimum around a few 100~$\mu$m.
\item Comparing our results with available observations showing a V-shaped polarisation spectrum implies that carbonaceous grains might indeed get destroyed in regions illuminated by a strong radiation field.
\item A change in the dust spectral index of silicate grains affects the steepness of the polarisation spectrum at long wavelengths.
\end{itemize}

Finally, we present a first polarisation spectrum from a self-consistent 3D molecular cloud simulation, which shows indications of a minimum in the polarisation spectrum around 350~$\mu$m. This shows the potential to further study polarisation spectra with state-of-the art 3D, MHD simulations. As a next step, we intend to include a more sophisticated treatment of the star formation process, the suggested carbon grain destruction and the emission of PAHs in the 3D, MHD model, which so far only have been accounted for in a simplified manner by using a lower dust-size limit of 5~nm. Furthermore, it might be of interest to repeat the current analysis with the new ``Astrodust'' model by \citet{Draine21}.

\section*{Acknowledgements}

The authors like to thank the anonymous referee for the comments which helped to significantly improve the paper. 
DS and SW acknowledge support of the Bonn-Cologne Graduate School, which is funded through the German Excellence Initiative as well as funding by the Deutsche Forschungsgemeinschaft (DFG) via the Collaborative Research Center SFB 956 ``Conditions and Impact of Star Formation'' (subprojects C5 and C6) and the SFB 1601 ``Habitats of massive stars across cosmic time'' (subprojects B1, B4 and B6).
Furthermore, the project is receiving funding from the programme ``Profilbildung 2020'', an initiative of the Ministry of Culture and Science of the State of Northrhine Westphalia. The sole responsibility for the content of this publication lies with the authors. T.B. acknowledges funding from the European Union H2020-MSCA-ITN-2019 under Grant Agreement no. 860470 (CHAMELEON).
The FLASH code used in this work was partly developed by the Flash Center for Computational Science at the University of Chicago. The authors acknowledge the Leibniz-Rechenzentrum Garching for providing computing time on SuperMUC via the project ``pr94du'' as well as the Gauss Centre for Supercomputing e.V. (www.gauss-centre.eu).

\section*{Data Availability}

The data underlying this article can be shared for selected scientific purposes after request to the corresponding author.




\bibliographystyle{mnras}
\bibliography{literature} 



\appendix

\section{Properly using two dust species in POLARIS}
\label{sec:appendixa}

POLARIS is able to differentiate between silicate and carbonaceous grains, giving tabulated data for each type, e.g. those found in the files \textsc{silicate\_oblate.dat} and \textsc{graphite\_oblate.dat} provided with the POLARIS distribution. As indicated in the POLARIS manual\footnote{\url{https://github.com/polaris-MCRT/POLARIS}}, the mixing ratio of both grain types can be set via
\begin{flalign}
 &  \textsc{$<$dust\_component$>$ "silicate\_oblate.dat" $f_\textrm{sil}$  \, .....} \, \nonumber \\
 &  \textsc{$<$dust\_component$>$ "graphite\_oblate.dat" $f_\textrm{carb}$ \, .....} \, \nonumber
\end{flalign}
in the POLARIS script file, where additional arguments are indicated by ``.....'' . The dust densities for silicate and carbonaceous grains will then be calculated from the total gas density $\rho_\textrm{tot}$, which is provided by the user via the input file, with the given fractions $f_\textrm{sil}$ and $f_\textrm{carb}$ and a dust-to-gas-ratio conversion factor (usually 0.01). The specification of this statement in the script file takes into account the different dust properties for the polarised radiative transfer calculation, e.g. it considers the fact that carbonaceous grains are not aligned as specified by default in the  \textsc{graphite\_oblate.dat} file. 

However, in the way outlined above, POLARIS will \textit{not} differentiate between the temperatures for silicate and carbon grains. Rather, when using the dust temperature calculation with the command ``\textsc{CMD\_TEMP}'', it will calculate a \textit{single}, averaged dust temperature for silicate and carbon grains together. This temperature can be size-dependent by including the statement
\begin{flalign}
 \textsc{$<$full\_dust\_temp$>$~1} \, \nonumber
\end{flalign} 
in the POLARIS script file, but even then for each grain size only one average dust temperature is calculated. However, as we have shown in Fig.~\ref{fig:polaris}, the differences between carbon and silicate grain temperatures are essential for the polarisation degree/spectrum. It is therefore indispensable to properly account for this difference as otherwise the resulting polarisation degree can be significantly falsified.

In order to calculate carbon and silicate grain temperatures separately, already the input file provided by the user has to be modified. Instead of containing only the total gas density $\rho_\textrm{tot}$, it must contain two density entries of the total density scaled with the two values of $f_i$, i.e. $f_\textrm{sil} \times \rho_\textrm{tot}$ and $f_\textrm{carb} \times \rho_\textrm{tot}$. Secondly, the two aforementioned lines in the POLARIS script file have to be modified to read
\begin{flalign}
 &  \textsc{$<$dust\_component id="1"$>$ "silicate\_oblate.dat" 1.  \, .....} \, \nonumber \\
 &  \textsc{$<$dust\_component id="2"$>$ "graphite\_oblate.dat" 1. \, .....} \,\, . \nonumber
\end{flalign}
We emphasise that it is essential to assure that values of \textsc{id}, "1" and "2", in these lines are identical to the ordering in the input file provided by the user. Furthermore, these lines need to be included for all steps, i.e. the calculation of the dust temperatures, the alignment, as well as the final radiative transfer. Doing so will provide the user with the grain-type specific temperatures as well as the resulting self-consistent polarisation degree.

\section{Supplemental figures}

\begin{figure*}
\centering
\includegraphics[width=0.9\linewidth]{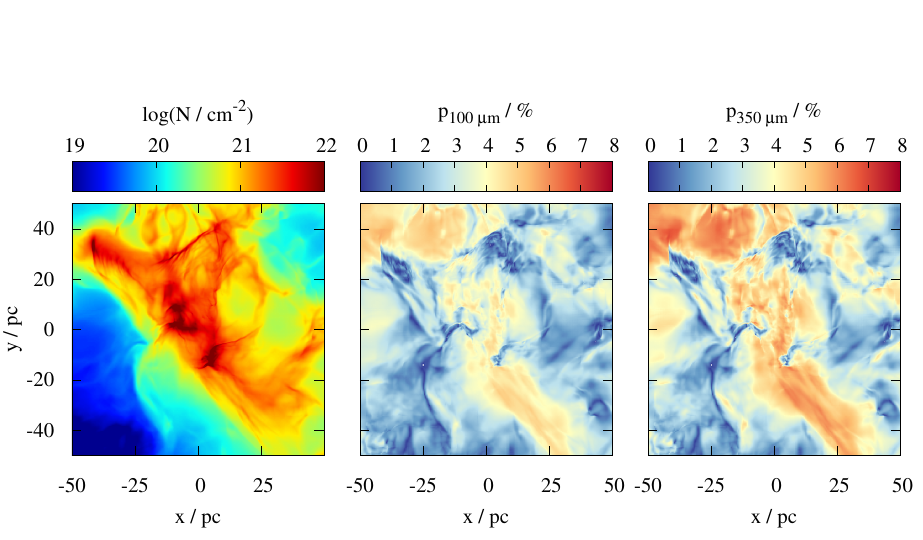} 
\caption{Maps of the column density (left) and the polarisation degree at 100~$\mu$m (middle) and 350~$\mu$m (right) for the 3D, MHD simulation of a molecular cloud. The cloud has a complex structure both in its column density and the polarisation degree. The polarisation degree at both wavelengths is qualitatively very similar, but consistently larger at 350~$\mu$m in accordance with the findings in Fig.~\ref{fig:ISRF}.}
\label{fig:map}
\end{figure*}

In Fig.~\ref{fig:map} we show the maps of the column density and the polarisation degree at 100 and 350~$\mu$m for the simulated molecular cloud analysed in Section~\ref{sec:3Dresult}, revealing its complex structure. The polarisation degree at both wavelengths is qualitatively very similar, but appears to be scaled up at 350~$\mu$m. This is in agreement with the findings in Fig.~\ref{fig:ISRF}, which show that the median polarisation degree at 350~$\mu$m is indeed higher than that at 100~$\mu$m.

%


\bsp	
\label{lastpage}
\end{document}